\documentclass{aims}

\usepackage{amsmath}
  \usepackage{paralist}
  \usepackage{graphics} 
  \usepackage{epsfig} 

 \usepackage[colorlinks=true]{hyperref}

\def\currentvolume{X}
 \def\currentissue{X}
  \def\currentyear{200X}
   \def\currentmonth{XX}
    \def\ppages{X--XX}

\newtheorem{theorem}{Theorem}[section]

\theoremstyle{definition}

\newcommand{\beq}{\begin{equation}}  
\newcommand{\eeq}{\end{equation}}  
\newcommand{\bea}{\begin{eqnarray}}  
\newcommand{\eea}{\end{eqnarray}}  
\newcommand\la{{\lambda}}  
  
\newcommand\al{{\alpha}}  
\newcommand\be{{\beta}}  
\newcommand\de{{\delta}}

\title[The Popowicz peakon system]
      {On the non-integrability of the Popowicz  peakon system}

\author[Andrew Hone and Michael Irle]{}


 \email{anwh@kent.ac.uk}
 \email{mvi3@kent.ac.uk}

\thanks{The second author is supported by an IMSAS studentship at the University of Kent.}

\begin{document}
\maketitle
 
\centerline{\scshape  Andrew N.W. Hone }
\medskip
\centerline{\scshape Michael V. Irle}

\medskip

{\footnotesize
 \centerline{Institute of Mathematics, Statistics \&
Actuarial Science} 
   \centerline{University of Kent,
Canterbury CT2~7NF, UK}  
} 



\bigskip

 \centerline{(Communicated by the associate editor name)}

\begin{abstract}
We consider a coupled system of Hamiltonian partial differential equations introduced by Popowicz, which has the appearance of a two-field coupling between the Camassa-Holm and Degasperis-Procesi equations. The latter equations 
are both known to be integrable, and admit peaked soliton (peakon) solutions with discontinuous derivatives at the peaks. A combination of a reciprocal transformation with 
Painlev\'e analysis provides strong evidence that the Popowicz system is non-integrable. Nevertheless, we are able to construct exact travelling wave solutions 
in terms of an elliptic integral, together with a degenerate travelling wave corresponding to a single peakon. We also describe the dynamics of $N$-peakon solutions, 
which is given in terms of a Hamiltonian system on a phase space of dimension $3N$. 
\end{abstract}

\section{Introduction}
The members of a one-parameter family of partial differential equations, 
namely  
\beq \label{bfam} 
m_t+um_x+bu_xm=0, \qquad m=u-u_{xx} 
\eeq 
with parameter %
$b$, 
have been studied recently. The case 
$b=2$ 
is the 
Camassa-Holm equation \cite{CH}, while 
$b=3$ 
is the Degasperis-Procesi equation 
\cite{DP}, 
and it is known that (with the possible exception of 
$b=0$) 
these are the 
only integrable cases \cite{miknov}, while all of of these equations 
(apart from 
$b=-1$) 
arise as a shallow water approximation to the 
Euler equations \cite{dgh}. All of the equations have at least one 
Hamiltonian structure \cite{HW}, this being given by 
\beq \label{bham} 
m_t = \mathcal{B}\frac{\de H}{\de m}, \quad  
\mathcal{B}=-b^2 m^{1-1/b}\partial_x m^{1/b}\hat{G}
 m^{1/b}\partial_x  m^{1-1/b} ,   
\eeq 
with 
$\hat{G}=(\partial_x-\partial_x^3)^{-1}$ 
and the 
Hamiltonian 
$H=(b-1)^{-1}\int m \, \mathrm{d} x$ 
for 
$b\neq 0,1$ 
(and the latter special cases admit a similar expression). 

One of the most interesting features of these 
equations is that their soliton solutions are not smooth, but rather the 
field 
$u$ 
has a discontinuous derivative at one or more peaks (hence the name 
{\it peakons}), while the corresponding field 
$m$ 
is measure valued. More precisely 
for the single peakon the solution has the form 
$$ 
u=c\, \exp (x-ct-x_0), \qquad  
\mathrm{with} \quad   
m=2c\, \de (x-ct-x_0)$$  
(with $x_0$ being an arbitrary constant), 
while 
$N$-peakon 
solutions are given by 
\beq \label{Npeaks} 
u=\sum_{j=1}^N p_j(t)\,\exp(-| x-q_j(t)|), \qquad m= 2\sum_{j=1}^Np_j(t)\de( x-q_j(t)), 
\eeq 
where the amplitudes 
$p_j(t)$ 
and peak positions 
$q_j(t)$ 
satisfy a Hamiltonian 
dynamical system for any 
$b$. 
For 
$b=2$ 
the 
$q_j$ 
and 
$p_j$ 
are canonically conjugate 
position and momentum variables for an integrable geodesic flow with the 
co-metric 
$g^{jk}=\exp (-|q_j-q_k|)$ 
\cite{CH}, and for 
$b=3$ 
the peakon motion is again integrable, being described by Hamilton's equations for a different   
Poisson structure \cite{dhh,dhh2}, but for arbitrary 
$b$ 
the $N$-peakon 
dynamics is unlikely to be integrable in general \cite{HH}. 

There is currently much interest in generalisations of the Camassa-Holm 
equations and its relatives. Qiao has found an integrable equation of this type 
with cubic nonlinearity \cite{qiao}, and another such example was 
discovered very recently by Vladimir Novikov \cite{volodya}; one of the authors spoke 
at the AIMS meeting in Arlington on this topic (for more details see 
\cite{wanghone2}). An important challenge is to understand the solutions 
of coupled equations with two or components, and in higher dimensions.
For example the EPDiff equation can be used to describe fluids in two or more spatial 
dimensions, as well as appearing in computational anatomy \cite{neuroepdiff}. 
Chen {\it et al.} 
found an integrable two-component analogue of the Camassa-Holm equation \cite{chen}, 
which also admits multi-peakon solutions \cite{falqui}. Popowicz has constructed 
another two-component Camassa-Holm equation using supersymmetry algebra \cite{Popsuper}. 

The purpose of this short note is to summarise some preliminary results that we 
have obtained on the two-component system given by  
\begin{equation}
\begin{array}{rcl} 
m_t + m_x (2u+v) + 3m (2u_x+v_x) & = & 0, \\
n_t + n_x (2u+v) + 2n (2u_x+v_x) & = & 0,  \\ 
 m=u-u_{xx}, &&  n=v-v_{xx}. 
\end{array}
\label{PopEq}
\end{equation}
%
This system can be considered as a coupling between the Camassa-Holm equation 
and the Degasperis-Procesi 
equation 
(corresponding to (\ref{bfam}) for $b=2,3$ respectively); it reduces to the former 
when $u=0$, and to the latter when $v=0$. The system (\ref{PopEq}) was obtained 
by Popowicz by taking a Dirac reduction of a three-field  
local Hamiltonian operator \cite{Pop}. By construction, this system 
has a (nonlocal) Hamiltonian structure, and due to the existence of 
conservation laws it was conjectured that it should be integrable 
(although no second Hamiltonian structure was found). 

After reviewing the Hamiltonian structure for it in the next section, in 
section 3 we perform a reciprocal transformation on the system (a nonlocal change 
of independent variables) which transforms 
it to a third order partial differential equation for a single scalar field. 
By applying Painlev\'e analysis of the singularities in solutions of 
the reciprocally transformed system, we find 
the presence of logarithmic branching, which is a strong indicator of non-integrability. 
Nevertheless, in section 4 we find that the system has exact travelling wave solutions given 
by an elliptic integral, as well as a degenerate travelling wave which is a peakon. In 
section 5 we present formulae for $N$-peakon solutions of (\ref{PopEq}), which   
are governed by Hamiltonian dynamics on a $3N$-dimensional phase space. The final section 
is devoted to some conclusions.

\section{Hamiltonian and Poisson structure}

Popowicz constructed the system  (\ref{PopEq}) from the 
Hamiltonian operator
\begin{equation}
\mathcal{Z} = - \left(
\begin{array}{cc}
9m^{2/3} \partial_x m^{1/3} \hat{G} m^{1/3} \partial_x m^{2/3} & 6m^{2/3} \partial_x m^{1/3} \hat{G} n^{1/2} \partial_X n^{1/2} \\
6m^{1/2} \partial_x n^{1/2} \hat{G} m^{1/3} \partial_x m^{2/3} & 4n^{1/2} \partial_x n^{1/2} \hat{G} n^{1/2} \partial_x n^{1/2}
\end{array}
\right),
\label{HamOp}
\end{equation}
where 
$\hat{G} = (\partial_x-\partial_x^3)^{-1}$. 
With 
the Hamiltonian
\begin{equation}
H_0 = \int (m+n) \, \mathrm{d}x, 
\label{Ham}
\end{equation}
the system can be written as 
$$
\left( \begin{array}{c} m_t \\ \\ n_t \end{array} \right) = 
\mathcal{Z} \left( \begin{array}{c} \frac{\delta H_0}{\delta m} 
\\ \\ \frac{\delta H_0}{\delta n} \end{array} \right)
\equiv \{ m,H_0 \} . 
$$
%
For $\mathcal{Z}$ as in (\ref{HamOp}), 
the Poisson bracket between two functionals $A,B$ is given by the standard formula
$$ \{A,B\} = \int 
\left(
\begin{array}{cc} \frac{\delta A}{\delta m(z)} & \frac{\delta A}{\delta n(z)} \end{array}
\right) \mathcal{Z} \left(
\begin{array}{c} \frac{\delta B}{\delta m(z)} \\ \frac{\delta B}{\delta n(z)} \end{array}
\right) \, \mathrm{d}z, 
$$
%
which is equivalent to specifying  the local Poisson brackets between the fields $m$ and $n$ as 
\begin{eqnarray} \label{mnpbs} 
\{ m(x),m(y) \} &=& m_x(x) m_x(y) G(x-y) \nonumber \\
& & + 3( m(x) m_x(y) - m_x(x) m(y) ) G'(x-y) \nonumber \\
& & - 9 m(x) m(y) G''(x-y), \nonumber \\
\{ m(x),n(y) \} &=& m_x(x) n_x(y) G(x-y) \nonumber \\ 
& & + ( 3 m(x) n_x(y) - 2 m_x(x) n(y) ) G'(x-y) \label{PBmn} \\
& & - 6 m(x) n(y) G''(x-y), \nonumber \\
\{ n(x),n(y) \} &=& n_x(x) n_x(y) G(x-y) \nonumber \\
& & + 2( n(x) n_x(y) - n_x(x) n(y) ) G'(x-y) \nonumber \\
& & - 4 n(x) n(y) G''(x-y), \nonumber
\end{eqnarray}
where 
\beq \label{greenfn} 
G(x) = \frac{1}{2} \mathrm{sgn}(x) \left( 1 - e^{- \vert x \vert} \right) 
\eeq  
is the Green's function of the operator $\hat{G}$. 
This $G$ satisfies the
functional equation 
$$ 
G'(\al )\Big(G(\be )+G(\gamma )\Big) +\mathrm{cyclic}=0  
\quad 
\mathrm{for}\,\, \al +\be +\gamma =0, 
$$ 
which is a sufficient condition for 
the operator (\ref{bham}) to satisfy the Jacobi identity; the general solution to 
the functional equation was found by 
Braden and Byatt-Smith in the appendix of \cite{HH}. 


It was observed by Popowicz that, apart from the Hamiltonian $H_0$, the 
system (\ref{PopEq}) has additional conserved quantities that 
can be written as 
\beq \label{cons} 
\begin{array}{rcl} 
H_1 & = & \int (nm^{-2/3})^\la m^{1/3}\,\mathrm{d}x, \\ 
H_2 & = & \int (-9n_x^2n^{-2}m^{-1/3}+12n_xm_xn^{-1}m^{-4/3}+-4m_x^2m^{-7/3}) 
\,(nm^{-2/3})^\la \, \mathrm{d}x,
\end{array} 
\eeq 
where in each case the parameter 
$\la$ 
is arbitrary. In \cite{Pop} it is remarked 
that, having three conserved quantities, the system is likely to be integrable. 
The existence of a mere three (or a few) conservation laws does not guarantee integrability, and a more 
precise requirement (or better, a definition of integrability) in infinite dimensions  
is that an integrable system should have infinitely many commuting symmetries \cite{miknov}. 
In fact, since they contain an arbitrary parameter, each of $H_1$ and $H_2$ provide 
infinitely many independent conservation laws for the system. However, a brief calculation 
shows that the gradient of each functional appearing in (\ref{cons}) is in the kernel of 
the Hamiltonian operator $\mathcal{Z}$ for all $\la$, so that all of these conserved 
quantities are Casimirs for the associated Poisson bracket. Hence, regardless of 
the choice of $\la$, neither $H_1$ nor 
$H_2$ can generate a non-trivial flow that commutes with the time evolution $\partial_t$.  

The fact that the combination 
$w=nm^{-2/3}$ 
appears in the conserved functionals (\ref{cons}) suggests 
that it is worthwhile to eliminate either 
$m$ or $n$ 
and use this as a dependent variable. 
Also, as noted by Popowicz, the 
conservation laws corresponding to 
$H_1$ 
are reminiscent of analogous ones for the Camassa-Holm/Degasperis-Procesi 
equations, which provide 
a reciprocal transformation to an equivalent system with different independent variables. 
We now make use of these observations.    

\section{Reciprocal transformation and Painlev\'{e} analysis} 

In order to eliminate 
$n$ 
we can rewrite 
(\ref{PopEq}) 
as 
\beq
\label{wsys} 
\begin{array}{rcl} 
(m^{1/3})_t & = & -(m^{1/3}C)_x, \\ 
w_t & = & -Cw_x, 
\end{array} 
\eeq 
with 
\beq 
\label{cdef} 
C=2u+v, \quad m=u-u_{xx}, \quad wm^{2/3}=v-v_{xx}. 
\eeq 
The first equation is in conservation form, 
and previous experience with the Degasperis-Procesi 
equation 
\cite{dhh} 
suggests taking the reciprocal transformation 
\beq
\label{rt} 
\begin{array}{rcl}
\mathrm{d}X & = & p\,\mathrm{d}x-C\,p\,\mathrm{d}t, \qquad p=m^{1/3}, \\ 
\mathrm{d}T & = & \mathrm{d}t, 
\end{array}
\eeq 
so that derivatives transform as $\partial_x = p\partial_X$, $\partial_t=\partial_T-Cp\partial_X$.

In terms of the new independent variables $X,T$ and the dependent variables $p,w$, the system 
(\ref{wsys}) becomes 
\beq\label{rtsys} 
\begin{array}{rcl} 
(p^{-1})_T & = & C_X, \\ 
w_T & = & 0, 
\end{array} 
\eeq 
and solving the latter two equations in (\ref{cdef}) for $u,v$ we can write 
$$
C=2u+v=2m+wm^{2/3}+(p\partial_X)^2(2u+v)=2p^3+wp^2+p(pC_X)_X. 
$$
Substituting back for $C_X$ from the first of (\ref{rtsys}) gives 
$C=p^3+wp^2+p(p(p^{-1})_T))_X$, and differentiating both sides of the latter 
with respect to $X$ and substituting for $C_X$ once more produces a single 
equation of third order for $p$, namely 
\beq\label{peq} 
p_{XXT}=\frac{p_Xp_{XT}}{p}+\frac{(1-p_X^2)p_T}{p^2}+2wpp_X+(w_X+6p_X)p^2.
\eeq 
From the second equation (\ref{rtsys}), the coefficient $w=w(X)$ is an 
arbitary $T$-independent 
function of $X$. It turns out that the presence of this arbitrary function 
provides an obstruction to integrability, from the point of 
view of the Painlev\'e analysis of the partial differential equation (\ref{peq}). 
It is also easy to calculate the images under this reciprocal 
transformation of the conserved densities corresponding 
to (\ref{cons}): the density for $H_1$ is transformed to $w^\la$, and that 
for $H_2$ becomes $-9w^{\la -2}w_X^2$, both of which are trivial (since 
$w$ is no longer a dynamical variable). 

To analyse the singularities of the equation (\ref{peq}) we apply the 
Weiss-Tabor-Carnevale test. The details of the analysis are almost identical to that 
for the equation obtained from (\ref{bfam}) by an analogous reciprocal transformation, 
so we will only give a brief description of the results and refer the 
reader to \cite{honepainleve} for details of a similar calculation. There are two types 
of local expansion around an arbitrary singular manifold  
$\phi = \phi (X,T)=0$ 
corresponding to singularities on the right hand side of equation (\ref{peq}). 
For simplicity we can take the Kruskal reduced ansatz $\phi = X-f(T)$ 
with $f$ an arbitrary function of $T$, and it is sufficient to take $w$ to 
be a non-zero constant.   For the first type of expansion, $p$ is regular as 
$\phi\to 0$, and we have $p= \pm \phi +\al_2\phi^2+\al_3\phi^3+\ldots$, 
where $\al_1(T)$ and $\al_3(T)$ are arbitrary. The resonances are at $-1,1,2$, 
corresponding to the arbitrariness of $f,\al_2,\al_3$ respectively, and  
all resonance conditions are satisfied, so that this defines the leading part of 
a local expansion that is analytic around $\phi = X-f(T)=0$. However, for the 
other type of expansion, we have $p=\sum_{j\geq -1}\be_j\phi^j$ which 
gives resonances $-1,2,3$.   The leading coefficient satisfies $\be_{-1}^2=-\dot{f}/2$, 
while at the next order $\be_0=-w/4$. The resonance $-1$ corresponds to $f$, and the 
resonance condition for the value $3$ (corresponding to $\be_2$) is satisfied 
automatically, but the resonance $2$ corresponding to $\be_1$ gives the additional 
condition 
\beq \label{reso} 
\dot{\be}_{-1}w=-\frac{\ddot{f}w}{4\be_{-1}} = 0  
\eeq 
which is not satisfied unless $w\equiv 0$ (since $f$ is arbitrary). When $w\equiv 0$ 
the equation (\ref{peq}) corresponds to the integrable Degasperis-Procesi equation  
(see \cite{dhh}; this is also clear from the fact that $n\equiv 0$ in that case). 
The failure of the resonance condition for non-vanishing $w$ means that 
the local expansion around a pole must be augmented with 
infinitely many terms in $\log\phi$, 
so that the Painlev\'e property does not hold. These logarithmic terms are an 
indicator that (\ref{peq}) is not integrable, and hence the system (\ref{PopEq}) 
cannot be.

\section{Travelling waves}

Travelling wave 
solutions of (\ref{PopEq}) arise 
by putting $$u(x,t)=U(s), \qquad v(x,t)=V(s), \quad \mathrm{with} 
\quad s=x-ct,$$
to get
\beq \label{trav} 
 M(2U+V-c)^3=K_1, \quad N(2U+V-c)^2=K_2, 
\eeq 
with $M=U-U''$, $N=V-V''$, where $K_1$, $K_2$ are constants, 
and $C(s)=2U+V=c-kM^{-1/3}$ where  $k=-K_1^{1/3}$. 
From this it is also apparent that $w=NM^{-2/3}=K_2K_1^{-2/3}=\ell=$constant. 
Comparing with (\ref{rt}) and (\ref{rtsys}) is clear that 
travelling waves of (\ref{PopEq}) are transformed to travelling waves 
$p(X,T)=P(S)$ of 
(\ref{peq}) moving with speed $k$, with    
$$P(S)=M^{1/3}(s), \quad S=X-kT, \quad \mathrm{d}S=M^{1/3}(s)\,\mathrm{d}s.$$ 
The ordinary differential equation for travelling waves of (\ref{peq}) 
(with constant $w=\ell$) can be integrated twice to yield 
$$ 
\left(\frac{dP}{dS}\right)^2= -\frac{2}{k}\Big(P^4+\ell P^3 + mP^2 +cP\Big)+1 \equiv 
Q(P), 
$$ 
for $k\neq 0$, where $m$ is another integration constant. This reduces to an 
elliptic integral of the first kind, 
$$S+\mathrm{const}=\int \frac{\mathrm{d}P}{\sqrt{Q(P)}},$$ so that $P$ is an 
elliptic function of $S$. 
Note that these travelling waves provide meromorphic solutions of (\ref{peq}), 
but this does not contradict the Painlev\'e analysis in the previous 
section, because for travelling waves the singular manifold is 
of the form $\phi = X-f(T)=S-S_0$ for constant $S_0$, so that 
$f(T)=kT+S_0$, implying $\ddot{f}=0$ which removes the obstruction 
to the Painlev\'e property in (\ref{reso}).     

In the original variable $s$, we have a third 
kind differential 
$$\mathrm{d}s =\frac{\mathrm{d}P}{P\sqrt{Q(P)}}$$ (so that 
$M(s)=P^3(S)$ has algebraic branch points as a function of $s$). 
For particular choices of constants, when the quartic $Q$ has a double 
root, the elliptic integral reduces to an elementary one in 
terms of hyperbolic functions, corresponding to smooth solitary wave solutions 
with the characteristic soliton shape.
  
Soliton-type travelling wave solutions must have a constant non-zero 
background, since the requirement   
that $U$ and $V$ tend to zero as $s \rightarrow \pm \infty$   
implies $K_1=0=K_2$  
in (\ref{trav}), hence $k=0$ and the above analysis does not apply. 
However, in this case we can have a weak solution of (\ref{trav}) 
which is the peakon solution
$$ u(x,t) = a \,e^{-\vert x - ct \vert}, \quad v(x,t) = b\, e^{-\vert x - ct \vert}, $$
and
$$ m(x,t) = 2a \delta (x-ct), \quad n(x,t) = 2b \delta (x-ct), $$
where $a$ is an arbitrary constant and $c=2a+b$ is the wave speed. In the next section 
we extend this to multi-peakon solutions. 

\section{Hamiltonian dynamics of peakons}

The $N$-peakon solutions have the appearance of a simple sum of $N$ single 
peakons but with both the  amplitudes and positions of the peaks being 
time-dependent, like so: 
\begin{equation}
u(x,t) = \sum_{j=1}^N a_j(t) e^{-\vert x - q_j(t) \vert}, \qquad v(x,t) = \sum_{j=1}^N b_j(t) e^{-\vert x - q_j(t) \vert},
\label{PPeak}
\end{equation}
where $a_j(t)$ and $b_j(t)$ are the amplitudes of the waves and $q_j(t)$ is the position of the peak of both waves. 
The main result is as follows.

\begin{theorem} \label{peakondynamics} 
The Popowicz system (\ref{PopEq}) admits $N$-peakon solutions of the form (\ref{PPeak}), 
where the amplitudes $a_j$, $b_j$ and positions $q_j$ satisfy the dynamical system 
\begin{eqnarray*}
\dot{a}_j &=& 2a_j \sum_{k=1}^N (2a_k+b_k) \, \mathrm{sgn}(q_j-q_k) \, e^{-\vert q_j-q_k \vert}, \\
\dot{b}_j &=& b_j \sum_{k=1}^N (2a_k+b_k) \, \mathrm{sgn}(q_j-q_k) \, e^{-\vert q_j-q_k \vert}, \\
\dot{q}_j &=& \sum_{k=1}^N (2a_k+b_k) \, e^{-\vert q_j-q_k \vert} 
\end{eqnarray*}
for $j=1,\ldots , N$. 
These equations are an Hamiltonian system 
$$ \dot{a}_j = \{ a_j,h \},\quad  
\dot{b}_j = \{ b_j,h \}, \quad   
\dot{q}_j = \{ q_j,h \}$$ 
with the Hamiltonian $h=2 \sum_{j=1}^N (a_j + b_j) $, and the Poisson bracket  
\begin{eqnarray*}
\{ a_j,a_k \} &=& 2 a_j a_k \mathrm{sgn}(q_j-q_k) e^{-\vert q_j - q_k \vert}, \\
\{ b_j,b_k \} &=& \frac{1}{2} b_j b_k \mathrm{sgn}(q_j-q_k) e^{-\vert q_j - q_k \vert}, \\
\{ q_j,q_k \} &=& \frac{1}{2} \mathrm{sgn}(q_j-q_k) \left( 1 - e^{-\vert q_j - q_k \vert} \right), \\
\{ q_j,a_k \} &=& a_k e^{-\vert q_j - q_k \vert}, \\
\{ q_j,b_k \} &=& \frac{1}{2} b_k e^{-\vert q_j - q_k \vert}, \\
\{ a_j,b_k \} &=& a_j b_k \mathrm{sgn}(q_j-q_k) e^{-\vert q_j - q_k \vert}.
\end{eqnarray*}
This Poisson bracket has $N$ Casimirs $\mathcal{C}_j= a_j/b_j^2$ 
for $j=1,\ldots , N$. 
\end{theorem} 

The proof of the above result, which will be presented elsewhere,
is based on integration of the equations (\ref{PopEq}) 
and the brackets (\ref{mnpbs}) against suitable test functions with support around each of the peaks. 
Here it is worth remarking that although the phase space has dimension $3N$, fixing the
values of the  $N$ Casimirs reduces the motion onto $2N$-dimensional symplectic leaves. 
Once this has been done, one can eliminate the $a_j$, say, and solve $2N$ equations 
for $b_j$, $q_j$.

\section{Concluding remarks}

The evidence from Painlev\'e analysis suggests very strongly that the system (\ref{PopEq}) 
is not integrable. 
This raises the question of whether the $N$-peakon system can be integrable for $N>1$. 
The Liouville-Arnold theorem  requires the existence of a further $N-1$ independent conserved quantities 
in involution (in addition to $h$ and the Casimirs $\mathcal{C}_j$ 
which satisfy $\{\mathcal{C}_j,F\}=0$ 
for any function $F$ on phase space). 
The first interesting case 
is the $2$-peakon problem, which requires just one additional conserved quantity. 
In fact, a direct calculation shows that the independent quantity 
$$ 
J=b_1b_2\Big( 1-\exp (-\vert q_1-q_2\vert )\Big)  
$$ 
is in involution with the Hamiltonian, $\{ J,h\}=0$, so 
that the $N=2$ peakon system is completely integrable.  

There is also the question 
of whether these peakons are stable solutions.  We propose to address these issues in future work.

\section*{Acknowledgements} AH is grateful to Jing Ping Wang for useful discussions, and thanks the 
organisers of the special session on integrable systems for inviting him to speak 
at the AIMS meeting in Arlington, Texas.

\medskip
Received September 2006; revised February 2007.

\medskip


\begin{thebibliography}{99}

\bibitem{CH}
     \newblock R. Camassa and D.D. Holm, 
     \newblock \emph{An integrable shallow water wave equation with peaked solitons},
     \newblock Phys. Rev. Lett., \textbf{71} (1993), 1661--1664.

\bibitem{chen}
\newblock M. Chen, S.-Q. Liu and Y. Zhang, 
\newblock A 2-component generalization of the Camassa-Holm equation and its solutions, 
\newblock Lett. Math. Phys. {\bf 75}, (2006), 1–15.  

\bibitem{DP}
     \newblock A. Degasperis and M. Procesi,
     \newblock \emph{Asymptotic integrability},
     \newblock Symmetry and Peturbation Theory (Rome) ed A. Degasperis and G. Gaeta (River Edge, NJ: World Scientific)
     \newblock (1998), 22--37.


\bibitem{dhh} 
\newblock A. Degasperis, D.D. Holm and A.N.W. Hone,  
\newblock \emph{A new integrable equation with peakon solutions}, 
\newblock Theor. Math. Phys. {\bf 133} (2002), 1461-72

\bibitem{dhh2} 
\newblock A. Degasperis, D.D. Holm and A.N.W. Hone,  
\newblock \emph{Integrable and non-integrable equations with peakons}, 
\newblock in ``Nonlinear Physics: Theory and Experiment. II'' (eds. M.J. Ablowitz, M. Boiti, F. Pempinelli and B. Prinari), 
\newblock World Scientific,  (2003), 37--43. 

\bibitem{dgh}
\newblock H.R. Dullin, G.A. Gottwald  and D.D. Holm,  
\newblock Physica D {\bf 190} (2004), 1--14. 

\bibitem{falqui} 
\newblock G. Falqui, 
\newblock \emph{On a Camassa-Holm type equation with two dependent variables}, 
\newblock J. Phys. A: Math. Gen. {\bf 39} (2006), 327--342.

\bibitem{honepainleve} 
\newblock A.N.W. Hone,
\newblock \emph{Painlev\'e tests, singularity structure and integrability},  
\newblock in ``Integrability'' (ed. A.V. Mikhailov), Springer Lecture Notes in Physics,  
\newblock 35 pages, Springer (2008) at press.  

\bibitem{neuroepdiff}
\newblock D.D. Holm, J.T. Ratnanather, A. Trouve and L. Younes, 
\newblock \emph{Soliton dynamics in computational anatomy}, 
\newblock NeuroImage {\bf 23} (2004), S170-S178. 
 
\bibitem{HH}
    \newblock D.D. Holm and  A.N.W. Hone,
    \newblock \emph{A class of equations with peakon and pulson solutions (with an appendix by Harry Braden and John Byatt-Smith)},
    \newblock Journal of Nonlinear Mathematical Physics, \textbf{12} (2005), 380--394.

\bibitem{HW}
     \newblock  A.N.W. Hone and J.P. Wang,
     \newblock \emph{Prolongation algebras and Hamiltonian operators for peakon equations}
     \newblock Inverse Problems, \textbf{19} (2003), 129--145.

\bibitem{wanghone2}
     \newblock  A.N.W. Hone and J.P. Wang,
\newblock \emph{Integrable peakon equations with cubic nonlinearity},  
\newblock J. Phys. A: Math. Theor., \textbf{41}
(2008), 372002.

\bibitem{miknov} 
\newblock A.V. Mikhailov and V.S. Novikov,
  \newblock \emph{Perturbative symmetry approach}, 
\newblock J. Phys. A: Math. Gen. {\bf 35} (2002),  4775--4790.

\bibitem{volodya} 
\newblock V.S. Novikov, 
\newblock \emph{Generalisations of the Camassa-Holm equation}, 
\newblock (2008), in preparation.

\bibitem{Popsuper}
    \newblock Z. Popowicz,
    \newblock  \emph{A 2-component or N=2 supersymmetric Camassa-Holm equation},
    \newblock  Phys. Lett. A \textbf{354} (2006), 110--114.


\bibitem{Pop}
    \newblock Z. Popowicz,
    \newblock  \emph{A two-component generalization of the Degasperis-Procesi equation},
    \newblock  J. Phys. A: Math. Gen. \textbf{39} (2006), 13717--13726.

\bibitem{qiao} 
\newblock Z. Qiao, 
\newblock \emph{A new integrable equation with cuspons and W/M-shape-peaks solitons},
 \newblock  J. Math. Phys {\bf 47}, (2006), 112701-9.


\end{thebibliography}
\end{document}